\documentclass{PoS}

\usepackage{amsmath}

\newcommand{\C}{{\mathbb C}}

\newcommand{\R}{{\mathbb R}}
\newcommand{\F}{{\mathbb F}}

\newcommand{\cK}{{\mathcal K}}
\newcommand{\cN}{{\mathcal N}}

\newcommand{\cO}{{\mathcal O}}

\newcommand{\cU}{{\mathcal U}}

 \newcommand{\sC}{{\sf{C}}}

\newcommand{\g}{ \gamma}
\newcommand{\s}{ \sigma }

\newcommand{\Dim}{\text{Dim}}
\newcommand{\Tr}{{\rm Tr}}
\newcommand{\diag}{{\rm Diag}}
\newcommand{\bee}{{\mathbf b}}
\newcommand{\wt}{ \widetilde }
\newcommand{\ses}{\setminus\!}

 \newcommand{ \cy}{ {\bf{c}} }
\def\ws{ \widetilde\sigma }
\def\wt{ \widetilde\tau }

\newcommand{\bea}{\begin{eqnarray}}
\newcommand{\eea}{\end{eqnarray}}
\newcommand{\be}{\begin{equation}}
\newcommand{\ee}{\end{equation}}
\newcommand{\beq}{\begin{equation}}
\newcommand{\eeq}{\end{equation}}

\newcommand{\la}{\langle}
\newcommand{\ra}{\rangle}

\def\ket#1{\left| #1\right\rangle}

\title{On the counting tensor model observables as $U(N)$ and $O(N)$ classical invariants}

\ShortTitle{Counting tensor model observables as $U(N)$ and $O(N)$ classical invariants}

\author{\speaker{Joseph Ben Geloun} %\thanks{A footnote may follow.}
\\
      LIPN, UMR CNRS 7030, Universit\'e Sorbonne Paris Nord, 93430 Villetaneuse, France \\
      International Chair in Mathematical Physics and Applications  \\
      ICMPA-UNESCO Chair, 072Bp50, Cotonou, Benin \\
        E-mail: \email{bengeloun@lipn.univ-paris13.fr}}

\abstract{Real or complex tensor model observables, the backbone of the tensor theory space, are classical  (unitary, orthogonal, symplectic) Lie group invariants. 
These observables represent as colored graphs, and that representation gives an handle to
study their combinatorial, topological and algebraic properties. 
We give here an overview of the symmetric group-theoretic formulation of the enumeration of unitary and orthogonal invariant observables 
which turns out to  bear a rich structure. 
From their counting formulae, one finds a correspondence with topological field theory on 2-cellular complexes that brings other interpretations of the same countings. Furthermore, tensor model observables span an algebra  that turns out to be semi-simple. Dealing with complex tensors, we discuss the representation theoretic base of the algebra making explicit its Wedderburn-Artin decomposition. 
The real case is more subtle as a base of its Wedderburn-Artin decomposition is yet unknown. }

\FullConference{Corfu Summer Institute 2019 "School and Workshops on Elementary Particle Physics and Gravity" (CORFU2019)\\
		31 August - 25 September 2019\\
		Corfù, Greece}

\begin{document}

\section{Introduction}

In any physical model,  interactions and observables play a fundamental role. 
For tensor models \cite{ambj3dqg}--\cite{razvanbook},  these observables  build from  the contractions of multidimensional arrays or tensors that transform
covariantly under the action of  some classical Lie groups. 
The most recent studies on tensor models over Lie groups consider   $U(N)$,  the  unitary group of order $N$, and $O(N)$, 
 the orthogonal group of order $N$. Note that much less is known about tensor models with $Sp(2N)$-invariants, $Sp(2N)$ being the (real or complex) symplectic group, see \cite{Carrozza:2018psc} and \cite{ABGD}.   Defined by contractions of tensors, the observables or  interactions of  tensor models   simply become  polynomial invariants of these classical Lie groups (for short we shall call them tensor invariants). Correlators in such models compute therefore, at the perturbative level, in terms of Feynman graphs involving tensor invariants as their vertices. 
 We can easily foresee that the quantum field theory calculations heavily rely on the diagrammatics and combinatorics of these objects. Hence,  a systematic
 combinatorial study of $U(N)$ and $O(N)$ classical invariants  has been 
 launched in the recent years bringing already a wealth of core results 
 \cite{BenRamg,Diaz:2017kub,deMelloKoch:2017bvv, BenSamj,Itoyama:2017wjb, 
Diaz:2018xzt, Diaz:2018zbg, Itoyama:2018but, Itoyama:2019oab, ABGD, Itoyama:2019uqv,Amburg:2019dnj} 
 
A preferred way of enumerating these invariants mainly rests on algebraic techniques of the symmetric groups. There are a lot of reasons why the use symmetric groups has become a
natural reflex and a dominating tool in the combinatorial study of tensor invariants. Indeed, the success of such studies has strongly benefited
from the expertise and techniques developed increasingly for matrices in recent years. 
Matrix models interconnect, in a nonexhaustive fashion, integrable models, 2D gravity, gauge theory, string theory and Riemannian geometry. The symmetric groups and their representation 
were established master tools to tame correlators and observables of matrix models, 
and thereby to understand the half-BPS sector of $\cN=4$ SYM  \cite{cjr,cr,BHR1,Kemp:2019log}.  This success emanates from importing 
Schur-Weyl duality as an instrument for grasping Gauge-String duality \cite{Ramgoolam:2008yr}. 
Furthermore, this algebraic implement highlights new correspondences between
countings in quantum field theory, matrix models, and string theory
\cite{Gal,FeynCount,doubcos,refcount,quivcal,Caputa:2013hr,Mattioli:2016eyp}. 
With all the results on matrix correlators, it may come as no surprise that a similar 
approach extends to tensor models and yields applications even beyond the realm of theoretical physics. In 
quantum information processing \cite{Ramgoolam:2018ceu} and  linguistics    \cite{Kartsaklis:2017lfq}\cite{Ramgoolam:2018xty} one also finds that matrix models and symmetric groups gather a renewed attention. 
 
 A few words about the interest for tensor models fall in line.  Tensor models
\cite{ambj3dqg,mmgravity,sasa1} were introduced as candidate theories for quantum gravity in higher dimensions \cite{Rivasseau:2013uca,Rivasseau:2016wvy,Delporte:2018iyf}. 
As expected, they were much more difficult to address
than matrix models \cite{Di Francesco:1993nw}, one of the few successful  candidate approaches  quantizing gravity in 2D. The theory of tensors became  only tractable after the 
inception of colored tensors \cite{color}. This class of tensor models supports a large $N$ expansion \cite{Gur4}  in a similar way that matrix models expand by the famous `t Hooft large $N$ limit  \cite{'tHooft:1973jz}. After, a variety of results came to light: the critical behavior of tensor models was uncovered
analytically  \cite{Bonzom:2011zz,Gurau:2013cbh,Gurau:2011kk,CarrooAdri}, entire  new families of quantum tensor field theories were found renormalizable both at the perturbative  
\cite{BenGeloun:2011rc}-\cite{BenGeloun:2017xbd} 
 and at the non-perturbative levels \cite{Eichhorn:2013isa}-\cite{Lahoche:2019vzy}. 
 More recently, the Sachdev-Ye-Kitaev (SYK) condensed matter model \cite{SYK,maldastan} proves to have the same diagrammatics as colored tensor models at large $N$  \cite{witten}. This opens a whole new avenue of research giving rise to an unprecedented interest of the community for tensor models  (see  \cite{BenGeloun:2017jbi}-\cite{Delporte:2020ddk} and references therein), hence the urge of finding new tools to understand
 them better.

The manipulation of symmetric groups and its representation theory in tensor models has
shed a new light on calculations, allowed one to discover genuine effects, and
bridged theories by uncovering new correspondences
 (bijections between different-looking objects). 
It also reveals hidden structures at the interface of three domains:  
combinatorics, algebra and topology/geometry. 
The exact enumeration complex tensor invariants connects to a topological field theory (TopFT)
that gives in return a geometrical interpretation of each observable as a branched cover of the 2-dimensional sphere \cite{BenRamg}. Tensor invariants may be regarded as the  generators of an algebra of observables with
interesting properties (semi-simplicity, orthogonal bases, gradation) \cite{BenSamj}. 
The two-point correlators of complex observables expand in these orthogonal bases. 
Many of these features of complex tensor models extend to the real case \cite{ABGD}. Once the enumeration of real tensor invariants sorted, their TopFT formulation finds a
bijection with the covers of the torus with defects, and their algebra possesses also orthogonal bases and is semi-simple. As another interesting by product of these analyses, 
 new integer sequences has been recorded in OEIS \cite{oeis} while also some sequences 
 therein got simplified. 

 This work delivers a summary of three contributions \cite{BenRamg}, \cite{BenSamj}, and \cite{ABGD}. 
We put in parallel the counting  of $U(N)$ tensor invariants and its corollaries \cite{BenRamg} \cite{BenSamj} and that of $O(N)$ tensor invariants \cite{ABGD}. The next section introduces our notation. Then 
section \ref{countU} focuses on the counting of complex tensor osbervables. 
The following section   \ref{countO} undertakes  the same analysis but for real tensors 
before a conclusion is drawn  in section  \ref{concl} with some perspectives of this work.

\section{Notation: complex and real tensors} 
 \label{not}
%---------------------------------------------------------------------------------------------------
The building blocks of the theory are complex and real tensors that we now introduce. 
This section follows \cite{BenRamg, ABGD, razvanbook}. 

Consider a field $\F=\R,\C$, and 
$d$ vector spaces $V_a$ over $\F$, $a=1,\dots, d$, $d \ge 2$, of respective dimensions $N_a$. 
We denote $\cU(N)$ either $U(N)$ or $O(N)$ depending on the complex
or real nature of the field $\F$, respectively.  Let $T$ be a multilinear map $\otimes_{a=1}^dV_a \to \F$ 
that we call a tensor of rank $d$ with components $T_{p_1, \cdots, p_d} \in \F$ (also called,  by abuse, the tensor itself), $p_a=1,\dots, N_a$. 
As the group $\otimes_{a=1}^d \cU (N_a)$ acts on $\otimes_{a=1}^dV_a$ in the fundamental representation,  $T$ transforms under the tensor product of $d$ fundamental representations of the  groups $\cU(N_a)$. Each group $\cU(N_a)$ acts  independently on a tensor index $p_a$ and we 
 write the component of the transformed tensor as: 
\bea
T^R_{p_1,\dots, p_d} = \sum_{q_k} R^{(1)}_{p_1q_1}  \dots R^{(d)}_{p_dq_d}  
T_{q_1,\dots, q_d} \,, \qquad R^{(a)} \in \cU(N_a)\;. 
\eea 
The observables of real or complex tensor models are the contractions of the tensors $T_{p_1,\dots, p_d} $ with respect to a trivial metric (however, the metric will be not trivial when dealing with the symplectic group, this case is not treated here; the interested reader is referred to  \cite{ABGD}). Such contractions are invariant under the action of $\otimes_{a=1}^d \cU (N_a)$ (note that there is subtlety
in the complex case as the action of $R\in  \cU (N_a)$ on a $T$ entails 
the action of $R^\dag \in  \cU (N_a)$ on a conjugate $\bar T$). One should pay attention on the fact that if we
call these contractions $\cU(N))$ invariants without making precise 
how many tensorial factors are involved, this is also an abuse. For real tensors, we must contract an even number of them
to obtain an invariant. For complex tensors, only contractions from conjugate tensor
components, $T$ and $\bar T$, are allowed. 
We can easily see that  tensor contractions generalize
matrix traces that are $\cU(N)$ invariants: $\Tr((MM^\dag)^{n})$, $n\ge 1$, for a matrice $M$ of size $N$. 
For this reason, we use the same notation $\Tr (\cdot)$ for tensors, hereafter. 
Another important feature of tensor contractions is that they encode  as  $d$-regular graphs with edge coloring with $d$ different colors, and each color at every vertex (representing each tensor)
represents an index of this tensor \cite{color,Bonzom:2011zz}. We will come back on this property  in the following section. 
Calling $\bee$ that representative colored graph, we denote the invariant  equivalently by 
 \bea
{\rm complex:} \qquad 
O_{\bee}(T, \bar T) &=&  \Tr_{\bee}( \bar T \cdot T \dots \bar T \cdot  T ) \crcr
{\rm  real:} \quad \qquad\; 
O_{\bee}(T) &=&  \Tr_{\bee}( T   \cdot T \dots   \cdot  T ) \;. 
\eea 
The dot means that some indices of the tensors get sum by other indices of other
tensors. The data of the graph $\bee$ is sufficient to determine according to which contraction pattern
the tensor indices are summed, and to each tensor contraction we have a unique graph associated
with it (up to isomorphism). The simplest non trivial tensor contraction denotes
$\Tr_{2}(T,\bar T)$ in the complex case and $\Tr_{2}(T)$ in the real case. They 
express as
\bea
{\rm complex:} \qquad  \Tr_{2}(T,\bar T) &= & \sum_{p_k=1, \dots, N_k} T_{p_1p_2 \dots p_d}
\bar T_{p_1p_2 \dots p_d}
\crcr
{\rm  real:} \quad \qquad\;   \Tr_{2}(T) &= & \sum_{p_k=1,  \dots,  N_k} (T_{p_1p_2 \dots p_d})^2 \;. 
\label{t2}
\eea
This simple quadratic invariant in the components of the tensor will play an important
role when we will define the Gaussian tensor field measure. 

In the following, most of the illustrations are made at fixed rank $d=3$ but it generally extends
in any $d$. We will also mostly reduce to $N_a= N$ for simplicity. 

\ 

\begin{figure}
\centering
\begin{minipage}[t]{.8\textwidth}
\centering
\includegraphics[angle=0, width=8cm, height=2cm]{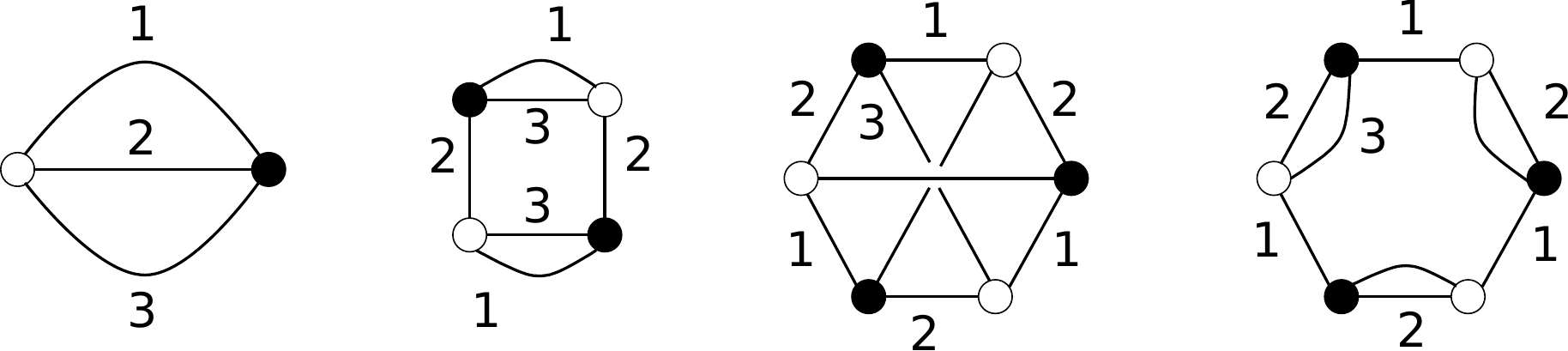}
\end{minipage}
\put(-250,13){$\bar T$}
\put(-290,13){$ T$}
\put(-280,-9){\small $\Tr_{2}(\bar T T)$}
\caption{$U(N)^{\otimes 3}$ tensor invariants.}
\label{fig:Uninv}
\end{figure}

\noindent{\bf Examples of unitary invariants.} Figure \ref{fig:Uninv}  illustrates contraction patterns 
for $U(N)^{\otimes 3}$ (complex) tensor invariants with 2 (one $T$ and one $\bar T$), 4, and 6 complex tensors. 
They are bipartite and colored graphs. $\bar T$ is represented by a black vertex, 
$T$ by a white one. The color $a=1,2,\dots,d$  of the edge is associated with the label $a$ of 
an index $p_a$ of the tensor.  Note that the graphs are bipartite. 

\begin{figure}
\centering
\begin{minipage}[t]{.8\textwidth}
\centering
\includegraphics[angle=0, width=8cm, height=2cm]{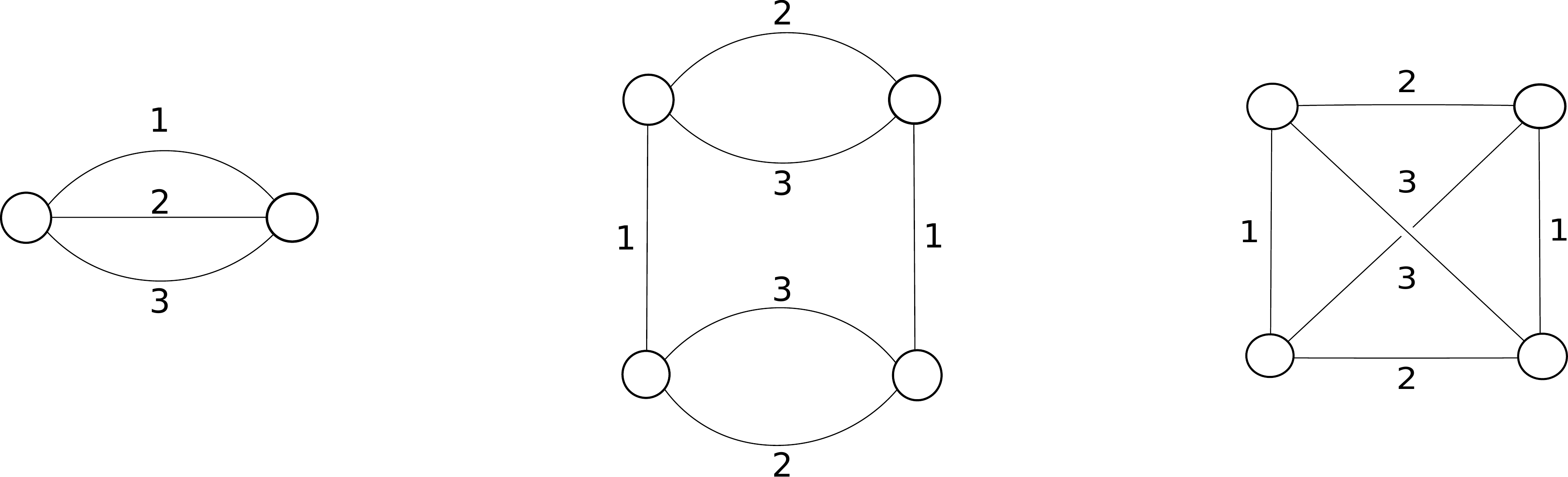}
\caption{$O(N)^{\otimes 3}$ tensor invariants.}
\label{fig:Oninv}
\end{minipage}
\put(-247,13){$T$}
\put(-287,13){$T$}
\put(-280,-9){\small $\Tr_{2}(T^2)$}
\end{figure}

\noindent{\bf Examples of orthogonal invariants.} 
Examples of $O(N)^{\otimes 3}$ tensor invariants are provided in Figure \ref{fig:Oninv}. 

\noindent{\bf Tensor field measure.}
Tensor contractions define the interactions of tensor models.  
We introduce a physical model through a partition function
(using abusively a single variable $T$ for both real and complex cases)
\bea
Z=\int d\nu(T)\exp(-S(T)) \, ,
\eea
where  $S(T)=\sum_{\bee}\lambda_{\bee} O_{\bee}(T)$ is a finite sum over 
some $\cU(N)$ invariants representing the model interactions, 
each associated  with a coupling $\lambda_{\bee}$.  The Gaussian tensor field measure
$d\nu(T)$ is given by 
\bea\label{GM}
{\rm complex:} \qquad &&
d \nu( T ) \equiv \prod_{i_k}
dT_{i_1i_2 \dots i_d}d\bar T_{i_1i_2 \dots i_d}\, e^{-\Tr_{2}(T,\bar T) }  \crcr
{\rm real:} \qquad &&
d \nu( T ) \equiv \prod_{i_k}
dT_{i_1i_2 \dots i_d }\, e^{- \Tr_{2}(T)}  \;,
\eea
with $\Tr_{2}(T,\bar T)$ and $\Tr_{2}(T)$ given by \eqref{t2}.
We will be interested in the moment of these measures called the 
correlators of the models:
\bea
\la{O_{\bee} (T) }  \ra =  \frac{ 1}{Z_0}
\int d \nu( T)  O_{\bee}(T) \; . 
\eea
with $Z_0 =\int d \nu( T)  $. 
The free propagator of the Gaussian measure corresponds to 
\beq\label{propag}
\la  T_{i_1 i_2 \dots i_d} T_{j_1 j_2 \dots j_d} \ra  = \frac{ 1}{Z_0} \int  d\nu(T) \,  T_{i_1 i_2 \dots i_d} T_{j_1 j_2  \dots j_d}  = \delta_{i_1j_1}\delta_{i_2j_2} \dots \delta_{i_d j_d} \,. 
\eeq
We will also discuss the mean values of two observables, called 2pt-function, 
\bea
\label{correlinit}
\la O_{\bee} (T)O_{\bee '} (T)  \ra = \frac{ 1}{Z_0} \int d\nu (T) O_b(T)O_{b'} (T)  \, .
\eea
The second correlator will be restricted to normal order
allowing only Wick contractions from $O_{\bee} (T) $ to $O_{\bee '} (T) $. 
Using the symmetric group formulation 
of the $\cU(N)$ invariants, we will reformulate \eqref{correlinit} and 
analyse the representation algebraic structure brought by the 2pt-function.

\section{Counting complex tensor model observables}
\label{countU}

Symmetric groups offer an elegant formulation of the counting 
of invariants based on the contractions of $n$ copies of 
tensors $T_{i_1, \cdots, i_d}$ and $n$ copies of the conjugate tensors $\bar T_{i_1, \cdots, i_d}$. 
This enumeration problem addresses as a TopFT with  permutation gauge group that 
we also discuss. 
Finally,  translating  the same counting in representation theory, 
we find a sum of  terms involving the famous Kronecker coefficient. 
This section summarizes  \cite{BenRamg} and \cite{BenSamj}. 

\subsection{Counting}
\label{cU}

Counting complex tensor invariants performs with the aid of a graphical representation
that we shall not refrain to explain. 
Each tensor $T_{p_1, \cdots, p_d}$ will be associated with a white vertex with $d$ exiting half-lines,
each of which is representing an index $p_a$, $a=1,\dots, d$.  The conjugate $\bar T_{p_1, \cdots, p_d}$ is 
represented by a black vertex with the same above principle for  its $d$ half-lines. Each index $p_a$ of any tensor is distinguished from the  others and no symmetry is assumed between the indices. For this reason, each contraction can only occur between indices $p_a$ of the same sub-label or color $a$, one belonging to a given $T$ and the other to a given $\bar T$. This contraction manifests at the level of the graph by connecting the two half-lines of index $a$
bewteen the two vertices associated with $T$ and $\bar T$.  

\begin{figure}[h]\begin{center}
     \begin{minipage}[t]{.8\textwidth}\centering
\includegraphics[angle=0, width=8cm, height=2.5cm]{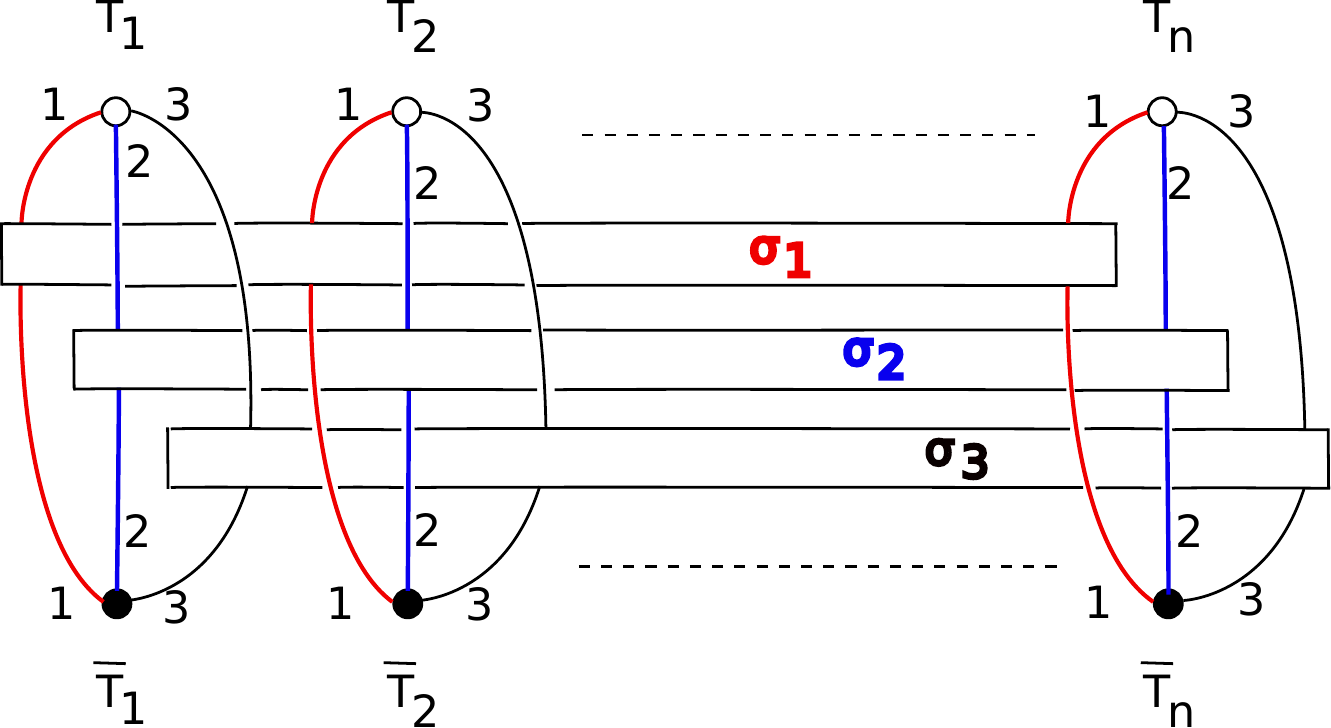}
\vspace{0.3cm}
\caption{ {\small  A tensor contraction identified as
permutation triple { $(\s_1,\s_2,\s_3)$}.}}
\label{countingC}
\end{minipage}
\end{center}
\end{figure}

We will concentrate on rank $d=3$,  as the general case $d$  recovers from this case. 
If one wishes to count  all the possible contractions 
between $n$ tensors  and $n$ conjugate tensors, this can be thought as counting all possible parings in the way given in Figure \ref{countingC}. 
Thus, we are enumerating permutation triples $( \s_1 , \s_2 , \s_3 ) \in  {S}_n \times {S}_n \times {S}_n$, 
$S_n$ being the symmetric group of $n$ elements, up to the equivalence 
 \bea 
( \s_1 , \s_2 , \s_3 ) \sim (\gamma_1  \s_1 \gamma_2 , \;
\gamma_1\, \s_2 \gamma_2 , \; \gamma_1 \s_3 \gamma_2  \,) \,, \quad 
\quad  \gamma_i \in {S}_n\,.
\eea 
This amounts to count elements of the double quotient   
 $ 
 {\diag} ( {S}_n ) \ses  ( {S}_n \times {S}_n \times {S}_n ) /  {\diag} ( {S}_n )\,.$ 
The enumeration of cosets relates to the counting of orbits of the left/right diagonal action. We therefore rely on Burnside's lemma 
and write, in terms of the fixed points of the same action described above, 
 \bea 
|H_1\ses G /\, H_2| =
 { 1 \over |H_1| |H_2| } 
\sum_{ h_1 \in H_1 } \sum_{ h_2 \in H_2 } \sum_{ g \in G } 
\delta   ( h_1 g  h_2 g^{-1} )  \,, 
\eea 
where the  $\delta$ is the usual delta function on the group, $\delta(\s)=1$, if $\s=id$, and $0$
otherwise. The  number of invariants expands as 
\bea
 Z_3(n) 
 &=&
  { 1 \over (n!)^2 } \sum_{ \sigma_{1,2,3} \in S_n}  \sum_{ \gamma_1, \g_2 \in S_n } 
    \delta ( \gamma_1 \sigma_1 \gamma_2^{-1 } \sigma_1^{-1} ) 
\delta ( \gamma_1 \sigma_2 \gamma_2^{-1} \sigma_2^{-1} ) 
 \delta ( \gamma_1 \sigma_3 \gamma_2^{-1 } \sigma_3^{-1} ) \crcr
 &=&
  \sum_{ p \,\vdash n }  {\rm Sym} ( p ) \,,   \qquad 
\qquad 
 {\rm Sym} (p):= \prod_{i=1}^n (i^{p_i})(p_i!)\; , 
 \label{blem}
 \eea
 where the sum is performed over all partitions $p$ of $n$, denoted $p \,\vdash n$. 
Programmed in Gap, and Mathematica, one obtains the sequence [OEIS: A110143 (isomorphism of graph coverings)]\cite{oeis}: 
\bea
1; 4; 11; 43; 161; 901; 5579; 43206; 378360; 3742738, ... 
\eea
The appendices of  \cite{BenRamg} store the programs used for this computation. 

The above generalizes for arbitrary rank $d$ using $d-$tuples of permutations $(\sigma_1, \dots, \sigma_d)$ $\in (S_n)^{\times d}$ equivalent under the diagonal 
action  $\diag (S_n)$ such that 
\be
(\sigma_1, \dots, \sigma_d) \sim (\gamma_1\sigma_1\gamma_2, \dots, \gamma_1\sigma_d\gamma_2)
\;.
\ee
We follow step by step the same procedure and in adapted notations and
 obtain the number of  rank $d$ tensor invariants made with $2n$ tensor fields as
\bea\label{ncopdtens} 
Z_d ( n ) = \sum_{ p \,\vdash n }  ( {\rm Sym} ( p ) )^{ d-2}  \qquad 
\qquad 
 {\rm Sym} (p):= \prod_{i=1}^n (i^{p_i})(p_i!)
\eea
Given $d$ and $n$, this number can be evaluated by a GAP or Mathematica program (see again appendices of \cite{BenRamg}). 
The link between the counting tensor invariants  and the counting of 
covers will become clearer when we will develop the permutation-TopFT formulation 
of the counting. Let us finally mention a word about connected invariants as the above counting includes all invariants connected and disconnected ones. To obtain connected invariants one should resort
by the so-called plethystic logarithm using the Moebius-mu function. The programs allowing to reach and sequences listing the numbers of connected invariants in the same reference.

\subsection{Topological Field Theory TopFT$_2$}
\label{topftU}
A quick look at \eqref{blem} shows a weighted sum of delta's functions. 
There is a simple physical construction, namely a
topological lattice gauge theory, where permutation groups play the role of 
gauge groups  \cite{quivcal} (henceforth called permutation-TopFT), that gives a sense of this expression. The topological invariance of this lattice construction illuminates the link between 
the counting of tensor invariants and the counting of branched covers of the 2-dimensional 
sphere. We start by the rank $d=3$ situation as its generalization naturally follows. 

\begin{figure}[h]
\begin{center}
 \begin{minipage}[t]{.8\textwidth}\centering
\includegraphics[angle=0, width=11cm, height=2.8cm]{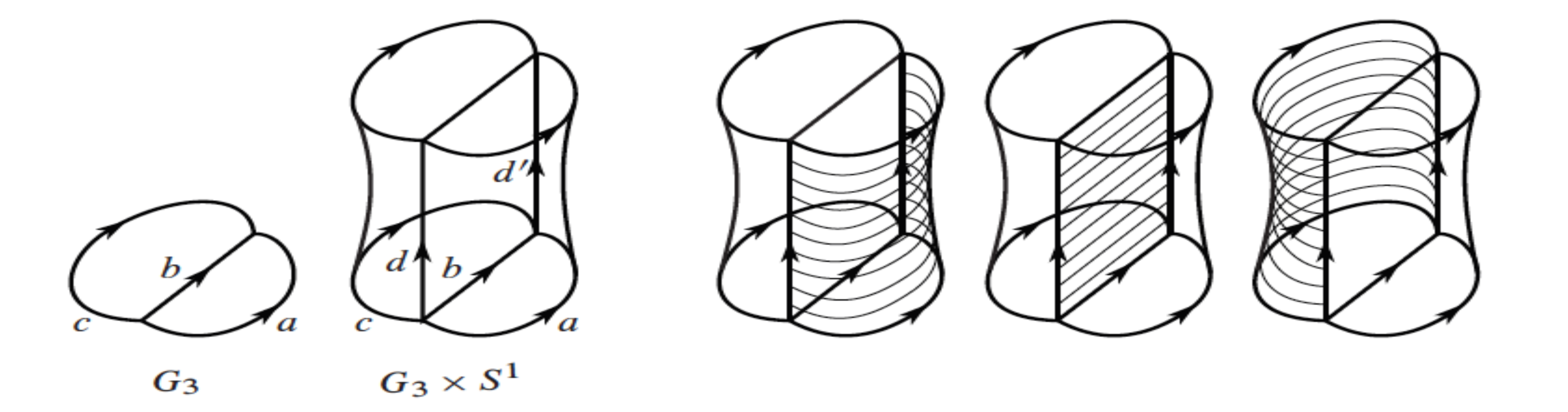}
\vspace{0.3cm}
\caption{ {\small The underlying lattice of the TopFT$_2$ related to $Z_3(n)$}.}
\label{latticeU}
\end{minipage}
\end{center}
\end{figure}

We look for a topological space leading to a permutation-TopFT whose
partition function corresponds to $Z_3(n)$ \eqref{blem}. 
Consider the graph $G_3$ in  Figure \ref{latticeU}, which has two vertices and 
three edges denoted by $a,b,$ and $c$.  Next consider $G_3 \times S^1$,  
which amounts to let evolve $G_3$ along a compactified time direction and then identifying the graph at the base of the Figure \ref{latticeU} with the one at the top. There are three (shaded) 2-cells of this cell-complex. 
To do $S_n$ permutation-TopFT (sometime as the lattice consists in gluing of 2-cells, we  further 
specify TopFT$_2$) on this complex, we assign 
$a \longrightarrow \s_1$, $b \longrightarrow \s_2$, $c \longrightarrow \s_3$ 
where the $\s_i \in S_n $. It is straighforward to check that the partition function of that
TopFT is precisely $Z_3(n)$. 

After some manipulations (gauge fixed one $\s_i$, say $\s_3$, set $\tau_i=\s_3^{-1}\s_i$, $i=1,2$,  rename $\g_2 \to \g$, and then introduce another variable $\tau_0 = (\tau_1\tau_2)^{-1}$), one arrives at:
 \bea 
 Z_3(n) =
 { 1 \over n! } \sum_{ \tau_0, \tau_{1},\tau_2 \in S_n}  \sum_{ \gamma \in S_n } 
\delta ( \gamma \tau_1 \gamma^{-1} \tau_1^{-1} ) 
 \delta ( \gamma \tau_2 \gamma^{-1 } \tau_2^{-1} )
 \delta ( \gamma \tau_0 \gamma^{-1 } \tau_0^{-1} )
 \delta(\tau_0 \tau_1 \tau_2) \;. 
\eea 
This formula therefore enumerates, according to the Burnside lemma, 
triples of permutations $ \tau_0 , \tau_1 , \tau_2 $ satisfying
\bea\label{tripleONE}  
\tau_0 \tau_1 \tau_2 = 1  \,. 
\eea
More precisely, it is counting equivalence classes of these triples under the conjugation 
equivalence by $\gamma  \in S_n$  : $ \tau_i \sim \gamma \tau_i \gamma^{-1} $. 
We recognize in (\ref{tripleONE})   the group generated by three generators subject 
to one relation, which is the fundamental group of the two-sphere, with three 
punctures.  Our counting function $Z_3 ( n )$  thus delivers the number of equivalence classes 
of branched covers of the 2-sphere, with 3-branch points, each equivalence class being counted 
once.  In two dimensions, branched covers are holomorphic maps and
so we expect that these permutation triples have rich properties: maps with three branch points  (often taken as $ 0,1,\infty $)  are called Belyi maps and are known to be definable over algebraic number fields \cite{Schneps}.  It turns out that  Belyi maps are also related
to large $N$ matrix models ribbon graphs  \cite{Gal}.  In fact, we can go deeper
in this puzzling correspondence between matrices and tensors (see section 8 of \cite{BenRamg}).  Matrix and tensor models 
have found a  clear connection: a Feynman diagram of a matrix model  relates to a state/observable of a rank-$3$ tensor model (with a certain weight).  This is a surprising feature reminiscent of dimensional reduction/uplift. 
This link has been deepen very recently by Amburg et al \cite{Amburg:2019dnj} as it generalizes at any $d$
(states in rank $d$ corresponds to diagrams in rank $d-1$). This of course deserves careful study as it may reveal important properties similar of that of  gauge/gravity duality for tensor models. 

Several other counting formulae for other types of  tensors have been digged out in \cite{BenRamg}, 
for instance, the case when the tensor is fully symmetric, or a color symmetrized counting. 
We shall not go futher by lack of place and invite the interested reader to have a look 
on this work.

\subsection{Algebras and representation theoretic bases}
\label{reptheor}

We will learn, in this subsection, another  piece of information about the counting
of complex tensor invariants under, this time,  a different light,  that of the representation theory of symmetric group $S_n$. 
For the basics of representation theory of the symmetric group, the reader may refer to the standard  textbook by Hammermesh \cite{Hammermesh}  and refer to \cite{BenSamj} for the calculations concerning tensors (appendices therein gather all what is needed in the paper). 

\ 

\noindent{\bf Mapping the counting in representations.}
The irreducible representations (irreps) of symmetric group $S_n$ are labeled by Young diagrams $R$, and these are  partitions of $n$: $R \vdash n$.   The Wigner matrices 
 $D^R_{ij}(\s)=\langle R,j |\s|R, i \rangle$  defines the real  matrix representation of  $\s \in S_n$  in the irrep  $R\vdash n$  of dimension $d(R)= n!/h(R)$, where $h(R)$ is the product of hook-lengths
 in the Young diagram $R$. The matrices  $D^R$ satisfy orthogonality properties and 
 convolute in well-known invariants of the representation theory of $S_n$.
 We will also need 
   the character $\chi^R(\cdot ) = \Tr(D^R (\cdot))$ of the representation $R$. 
   Any $\delta$ on the group expands as $\delta (\cdot)= \sum_{R \vdash n}[d(R)/n!]\chi^R(\cdot)$. 
  In a nutshell, the following  identities hold: 
 \bea
&&
{\rm orthogonality }: \quad \sum_{\s \in S_n} D^R_{ij}(\s)  D^S_{kl}(\s)
 = \frac{n!}{d(R)} \,\delta^{RS}\,\delta_{ik}\delta_{jl}\;;
\cr\cr
&&  {\rm reality}: \quad  D^{R}_{ij}(\s^{-1}) = D^R_{ji}(\s)\,; \\
 &&
 {\rm Clebsch-Gordan}: \quad
\sum_{\s\in S_n} D^{R_1}_{i_1j_1}(\s)D^{R_2}_{i_2j_2}(\s)D^{R_3}_{i_3j_3}(\s)
 = \frac{n!}{d(R_3)}\sum_{\tau}
C^{R_1,R_2;R_3,\tau}_{i_1,i_2;i_3}
C^{R_1,R_2;R_3,\tau}_{j_1,j_2;j_3} \nonumber \cr\cr
&&
\tau \in [\![1, \sC(R_1,R_2,R_3) ]\!] \,, 
\eea 
where the symbol 
\be 
\sC ( R_1 , R_2 , R_3 ) = \frac{1}{n!}
\sum_{\s \in S_n}  \chi^{ R_1  } ( \sigma )  \chi^{ R_2} ( \sigma) 
\chi^{ R_3} ( \sigma)   
\ee 
stands for the so-called Kronecker coefficient. The Kronecker coefficient is the 
multiplicity of the one-dimensional (trivial) representation in the tensor product
   $ R_1 \otimes R_2 \otimes R_3$. 

Expanding \eqref{blem} in irreps, we have by a small calculation 
\bea 
 Z_3(n) &=& 
  { 1 \over (n!)^2 } \sum_{ \sigma_i \in S_n}  \sum_{ \g_1, \g_2 \in S_n }  \delta ( \gamma_1 \sigma_1 \gamma_2^{-1 } \sigma_1^{-1} ) 
\delta ( \gamma_1 \sigma_2 \gamma_2^{-1} \sigma_2^{-1} ) 
 \delta ( \gamma_1 \sigma_3 \gamma_2^{-1 } \sigma_3^{-1} ) 
  \cr  
& =&  { 1 \over (n!)^2 } \sum_{ \g_i  \in S_n} \sum_{ R_i \vdash n  } \chi^{ R_1  } ( \g_1 )  \chi^{ R_1} ( \g_2 ) \chi^{ R_2  } ( \g_1 )  \chi^{ R_2} ( \g_2 ) \chi^{ R_3  } ( \g_1 )  \chi^{ R_3} ( \g_2 )  \cr 
&=&  \sum_{ R_1 , R_2 , R_3 \vdash n} (  \sC ( R_1 , R_2 , R_3) )^2   \; . 
\label{coefKon3}
\eea 
The counting of rank $d=3$ observables equates to a sum of square of Kronecker 
coefficients (shortly called at times Kroneckers). The above identity is of a certain interest as it is a longstanding open problem to 
give a combinatorial sense of $ \sC ( R_1 , R_2 , R_3)$  thereby ensuring its positivity (see Problem 10 in  \cite{Stanleyproblems};  
keeping in mind  the way that Littlewood-Richardson coefficient have found a combinatorial description).
Furthermore, this very coefficient attracks today a lot of attention in the
theoretical computer science community, in particular in computational complexity theory \cite{iken1, Blasiak} as it turns out to be the object of interest for understanding  a geometrical version of the famous problem  $P$ vs $NP$. 
We have given above a combinatorial interpretation of a sum of squares
of Kroneckers as the number of $3$-regular  edge-colored bipartite   graphs. 
A question arises: is there  a refinement of the counting of rank 3 observables that  boils down to a single coefficient? 
This should be investigated. Finally, to change group invariance, namely from $U(N)$ to $O(N)$, as we will see in the next section, also impacts \eqref{coefKon3}.
 
 The relation \eqref{coefKon3} generalizes  in any rank $d$: 
  $Z_d(n) = \sum_{ R_1 , R_2 , R_3 \vdash n} (  \sC ( R_1 , R_2 , \dots , R_d) )^2$, where
  $\sC ( R_1 , R_2 , \dots , R_d)$ is not the Kronecker coefficient (as this name is reserved for 3 irreps)
  but a generalized version of it:  it counts the multiplicity of the trivial representation in the tensor product
  of all $R_i$ listed.

\

\noindent{\bf $\cK_d(n)$, the double coset graph algebra.}
We now introduce an algebraic structure on observables.  Such an algebra has been 
very fruitful in matrix models since it allows one to uncover computable sectors. 
Consider the group algebra { $ \C ( S_n )$}, i.e. the space of linear combinations
 which  read $a = \sum_{\s\in S_n} \lambda_\s \s,$ with $\lambda_\s \in \C$. 
 As customary, we start by $d=3$ for simplicity.

 In the double coset formulation, we counted  the orbits 
 \bea \label{orbU}
( \s_1 , \s_2,\s_3  ) \sim  ( \g_1 \s_1 \g_2 , \g_1 \s_2 \g_3,  \g_1 \s_3 \g_2 )  \,.
\eea 
We  now want to  embed these orbits in $ \C ( S_n )^{ \otimes 3}$. For this purpose, 
define  $\cK_3(n) \subset  \C ( S_n )^{ \otimes 3}$  as the vector space over  $\C$ 
such that:
\be
\cK_3(n) = {\rm Span}_{\C}\Big\{    
\sum_{ \gamma_1, \gamma_2 \in S_n } \g_1 \s_1 \g_2 \otimes \g_1 \s_2 \g_2\otimes \g_1 \s_3 \g_2 , \; \s_1, \s_2, \s_3 \in S_n 
\Big\} \,. 
\label{graphbasis}
\ee 
By construction,  it is an obvious fact  that $\dim_\C \cK_3(n)= Z_3(n)$, as a
generator of $ \cK_3(n)$ is precisely (the sum of all elements of) an orbit of the left/right diagonal group action \eqref{orbU}. 

The vector space $\cK_3(n)$ has some properties that we now examine.
Introducing a minor change by a convenient normalization, any basis element of $\cK_3(n)$
expresses as  
\be
A_{\s_1,\s_2,\s_3} = \frac{1}{(n!)^2}\sum_{\g_1,\g_2 \in S_n} \g_1 \sigma_1 \gamma_2  \otimes \gamma_1 \sigma_2 \gamma_2 \otimes \gamma_1  \sigma_3 \gamma_2  \,. 
\ee
One should keep in mind that $A_{\s_1,\s_2,\s_3}$ corresponds to a graph (observable). 
We can multiply two such elements and obtain: 
 \bea
A_{\s_1,\s_2,\s_3} A_{\s_4,\s_5,\s_6}  &=&
  \frac{1}{n!}
\sum_{\tau\in S_n}
A_{\sigma_1 \tau \sigma_4,\, \sigma_2 \tau \sigma_5,\,\sigma_3 \tau \sigma_6}  \,.
\label{AB}
\eea
That means that the product is stable in $\cK_3(n)$, and therefore it makes it an algebra.
From such a property, we infer that graphs multiply in  $\cK_3(n)$, which henceforth can be called
a graph algebra.  $\cK_3(n=1)\equiv  \{e\}$ is trivial. The work \cite{BenSamj}  provides the  multiplication tables for $\cK_3(n=2)$ (isomorphic to $\C(S_n)\otimes \C(S_n)$) and $\cK_3(n=3)$,
both are them are commutative. 

It is also not difficult to realize that the product of $\cK_3(n)$ is associative and admits a unit, 
namely the orbit containing $(id,id,id)$. 
We conclude that $\cK_3(n)$ is an associative unital subalgebra of  $\C ( S_n )^{ \otimes 3}$. 
Moreover, $\cK_3(n)$   is semi-simple with the nondegenerate pairing [Prop. 2, \cite{BenSamj}]
\be
\delta_3 (  \otimes_{i=1}^3 \s_i ; \otimes_{i=1}^3 \s'_i ) = 
\prod_{i=1}^3 \delta (\s_i\s'^{-1}_i)  \,.
\label{delta3}
\ee

All the above properties extend at any rank $d\ge 3$: $\cK_d(n)$ is an associative unital 
semi-simple subalgebra of $\C ( S_n )^{\otimes d}$.  It is an interesting theme
of research to  determine the  structure coefficients of the graph algebra $\cK_d(n)$ 
and check if it has some known  isomorphism class. 
The semi-simplicity property of an algebra entails, by the Wedderburn-Artin theorem, 
that the algebra admits a  decomposition as a direct sum of matrix subalgebras. 
$\cK_d(n)$ therefore admits a Wedderburn-Artin decomposition. In fact, 
from the identity  $Z_d(n) = \sum_{ R_1 , R_2 , R_3 \vdash n} (  \sC ( R_1 , R_2 , \dots , R_d) )^2$, 
the dimension of each matrix subalgebra read-off as the square of the multiplicity $(  \sC ( R_1 , R_2 , \dots , R_d) )^2$.
In the following, we identify a basis of this matrix subalgebras in rank $d=3$. 

\

\noindent{\bf $\cK_3(n)$ as a centralizer algebra.} There is another formulation 
of the counting of rank $d$ observables by gauge fixing one $\s_i$. 
In rank $d=3$, we could choose to fix $\s_1$, 
$( \sigma_1 , \sigma_2 , \sigma_3 ) \rightarrow ( 1 , \sigma_1^{-1} \sigma_2 , \sigma_1^{-1} \sigma_3 ) \equiv ( 1 , \tau_1 , \tau_2 )$. The action of $ \g_2 = \g  $ reduces to  $ ( \tau_1 , \tau_2 )  $ as 
\bea 
( \tau_1 , \tau_2  ) \sim  ( \g \tau_1 \g^{-1} , \g \tau_2 \g^{-1} ) 
\eea
In the same way as previously done, we can span the algebra $\cK'_3(n)$
over the orbits of  couples  $ \tau_1 \otimes  \tau_2  $. We show that 
 it is a subalgebra of the group algebra $ \C ( S_n ) \otimes \C ( S_n )$ 
 which is invariant under conjugation by  the diagonally embedded   
$ S_n $. In such a  setting, we call $\cK'_3(n)$, a permutation centralizer algebra.
All properties of the double coset formulation recover here as well, simply because 
$\cK'_3(n)$ and $\cK_3(n)$  are isomorphic.  

\ 

\noindent{\bf $\cK_3(n) $ decomposed in matrix blocks.}
We investigate a base of $\cK_3(d)$ which makes explicit the Wedderburn-Artin 
decomposition.  Start by the Fourier basis of  $\C(S_n)$ defined as 
\be\label{fouU}
Q^R_{ ij}   =   { \kappa_R \over n! } \sum_{ \sigma \in S_n } D^R_{ ij} ( \sigma ) \sigma  \,, 
\ee
where $\kappa_R$ is a normalization factor.  Let   $\rho_L$ and $\rho_R$ denote  the left and right multiplications on $\C(S_n)^{\otimes 3}$, respectively. Then, we introduce the following 
convolution 
\bea
&&
 Q^{R_1,R_2,R_3}_{\tau,\tau'}  = 
\sum_{i_l, j_l,k} C^{R_1,R_2;R_3, \tau}_{i_1,i_2;i_3} C^{R_1,R_2;R_3, \tau'}_{j_1,j_2;j_3}
\sum_{\s_1,\s_2} \rho_L(\s_1)\rho_R(\s_2)
Q^{ R_1}_{ i_1 j_1 }\otimes 
Q^{ R_2}_{ i_2 j_2 } \otimes Q^{ R_3}_{ i_3 j_3 }  \crcr
&&
= \kappa_{R,S,T} \sum_{\s_l \in S_n} \sum_{i_l,j_l} 
C^{R , S ; T , \tau_1  }_{ i_1 , i_2 ; i_3 } C^{R , S ; T , \tau_2  }_{ j_1 , j_2 ; j_3 } 
D^{R}_{ i_1 , j_1 } ( \sigma_1 )
 D^{S}_{ i_2, j_2 } ( \sigma_2 ) D^{T}_{ i_3 , j_3 } ( \sigma_3 )
\sigma_1 \otimes \sigma_2 \otimes \sigma_3 \,. 
\eea
To check that the set $\{Q^{R_1,R_2,R_3}_{\tau,\tau'}\}$ forms an orthogonal matrix base of $\cK_3(n)$ is the next task. Provided a rightful choice of $\kappa_{R}$,  these elements multiply  like  matrices: 
\be
Q^{R,S,T}_{\tau_1,\tau_2}Q^{R',S',T'}_{\tau_2',\tau_3}   = 
\delta^{RR'} \delta^{SS'}  \delta^{TT'}\delta_{\tau_2 \tau_2'}  Q^{R,S,T}_{\tau_1,\tau_3}  \,.  
\ee
 Then observe that, at fixed $[R_1,R_2,R_3]$, $Q^{R_1,R_2,R_3}_{\tau,\tau'}$ is  matrix with $\sC(R_1,R_2,R_3)^2$ entries.  The fact they are orthogonal with respect to the pairing $\delta_3(\cdot)$ \eqref{delta3} can  be also derived in a direct manner.  Thus, we can infer that this is the Wedderburn-Artin basis for  $\cK_3(n)$. 
 
The scrutiny of $\cK_3(n)$ reports other results concerning   the existence of overcomplete
 bases, a nontrivial center $\mathcal{Z}(\cK_{3}(n))$ with base elements 
 given by $P^{R,S,T} = \sum_\tau Q^{R,S,T}_{\tau,\tau}$.  $P^{R,S,T}$  counts the number of
 nonvanishing Kronecker coefficients, another property which might be useful for 
 computational complexity theory. In the next section, we engage the quantum theory 
 and underline a few properties induced by the existence of orthogonal bases. 
 
 \

\noindent{\bf Correlators.} At rank $d=3$, we focus the Gaussian model \eqref{GM}
and aim at calculating 1pt- and 2pt-correlators. Especially in this section, we use $N_a=N$, $a=1,\dots, d$. 
Let  $ \cO_{ \bee } =  \cO_{ \s_1 , \s_2 , \s_3 }$ 
be an observable determined by a triple $(\s_1 , \s_2 , \s_3)$. By the Wick theorem, 
we obtain 
\bea 
\la \cO_{ \s_1 , \s_2 , \s_3 }  \ra
 = \sum_{\mu \in S_n} N^{\cy(\mu \s_1)+\cy(\mu \s_2)+\cy(\mu \s_3)} 
\eea 
$\cy(\alpha)$ is the number of cycles of $\alpha$. We map this expression in the Fourier 
components by contracting it with characters, at fixed $(S_1,S_2,S_3)$: 
\bea 
 \langle \cO_{ S_1 , S_2 , S_3  }  \rangle  
&=& { 1 \over (  n!)^3 }  \sum_{ \sigma_l \in S_n }
\chi^{ S_1} ( \sigma_1 ) \chi^{S_2} ( \sigma_2 ) \chi^{S_3}  ( \sigma_3 )
\langle \cO_{\sigma_1,\s_2,\s_3 }  \rangle  \crcr
&=& \frac{1}{(n!)^2}
\Big[\prod_{l=1}^3 f_N(S_l)\Big]~ \sC(S_1,S_2,S_3) \,, 
\label{O1sss}
\eea
where $f_N(R)$ is the product of  box weights after filling $R$. 
Hence, the correlators $\langle \cO_{ S_1 , S_2 , S_3}  \rangle $   are proportional
to the Kronecker coefficients. A similar expression hold for rank $d$ 1pt-function,
where the Kronecker is replaced by the coefficient $\sC(S_1,S_2,\dots ,S_d)$. 
Hence using software programs (in particular Sage has  a useful and efficient package to compute
Kroneckers), we can compute explicitly this sector of the Gaussian tensor model, and  
that is quite remarkable. 

The last thing that we wish to sketch is that normal ordered 2pt-functions
evaluate wtih the Wedderburn-Artin base $\{Q^{R,S,T}_{\tau,\tau'}\}$. 
We have (see equation (97) in  \cite{BenSamj}): 
\be \label{oco}
\langle \cO_{ \sigma_1 , \sigma_2 , \sigma_3 } \cO_{ \tau_1 , \tau_2 , \tau_3 } \rangle 
= \sum_{ \gamma_1 , \gamma_2 } N^{ 3 n } \delta _3[(  \sigma_1 \otimes \sigma_2 \otimes \sigma_3 ) \gamma_1^{ \otimes 3 }   (  \tau_1 \otimes \tau_2 \otimes \tau_3 ) \gamma_2^{ \otimes 3 }
( \Omega_1 \otimes \Omega_2 \otimes \Omega_3 ) ] \,, 
\ee
where  $\Omega_i =\sum_{\alpha_i\in S_n}N^{\cy(\alpha_i)-n}\alpha_i$  are shown to be central elements of $ \C ( S_n)$. Now, we introduce the Fourier (or representation basis ) of the observables: 
\bea 
\cO^{ R , S , T }_{ \tau_1 , \tau_2}= \sum_{ \sigma_1 , \sigma_2 , \sigma_3 } 
\delta_3  ( Q^{ R , S , T}_{ \tau_1 , \tau_2 } \sigma_1^{-1} \otimes \sigma_2^{-1} \otimes \sigma_3^{-1} ) 
\cO_{ \sigma_1 , \sigma_2 , \sigma_3 }  \,. 
\eea
A  calculation leads to 
\bea
\langle \cO^{ R_1 , S_1 , T_1 }_{ \tau_1 , \tau_1' } \overline { \cO^{ R_2 , S_2 , T_2 }_{ \tau_2 , \tau_2' } }  \rangle 
=  c(R_1,S_1,T_1) \delta_{ R_1 , R_2 } \delta_{ S_1 , S_2 } \delta_{ T_1 , T_2 } 
\delta_{ \tau_1 ,\tau_2 }\delta_{ \tau_1'  ,\tau_2' }  \,, 
\eea
with $c(R_1,S_1,T_1)$ a constant depending on representation indices. 
This demonstrates that  $\{\cO^{R,S,T}_{\tau,\tau'}\}$ forms
an orthogonal  basis for Gaussian normal ordered correlators 
arising directly from the $Q^{ R , S , T }_{ \tau , \tau'} $, which are the 
representation theoretic  base elements of $\cK_3(n)$. We emphasize that 
more results and the proofs of above identities are available in \cite{BenSamj}. Concerning correlators,  we obtain new correspondences with cover countings in a TopFT$_2$. On another side, a thorough exploration of the color symmetrized counting exhibit a graded algebra structure with grade labeled by the irreps of $S_d$, the group of permutation of the colors $ [\![1, d]\!] $. 

%----------------------------------------------------------------------------------------
\section{Counting real tensor model observables}
\label{countO}

This section deals with  $O(N)$ invariants for real tensor models. 
We adopt the same  methodology of the previous section. 
This section reports some of the  main results of \cite{ABGD}. 

\subsection{Counting}
\label{cO}
Tensors in this part are real and still denoted  $T_{p_1,\dots, p_d}$. 
Seeking a graphical representation for the contractions of tensors,  we 
use the same scheme as in subsection \ref{cU}. 

\begin{figure}[h]\begin{center}
     \begin{minipage}[t]{.8\textwidth}\centering
\includegraphics[angle=0, width=8cm, height=3cm]{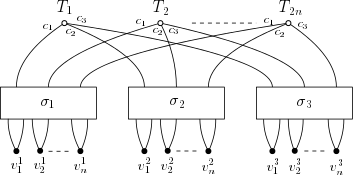}
\caption{Diagrammatics for real tensor contractions.}
\label{coloredgraphs}
\end{minipage}
\end{center}
\end{figure}

Orthogonal invariants bijectively correspond to  $d$-regular colored graphs (no bipartiteness)  \cite{CarrooAdri}. Indeed, at the contrary of the previous section, there is no objection that $T$ contracts its indices with another $T$. However, the bipartiteness is still useful to achieve
a counting. A way to restore this property consists in the insertion of another type of vertex
of valence 2, called black vertex, on each edge of the graph. We keep the initial vertices associated with $T$ as white. Consider the contraction of $2n$ tensors or white vertices and complete the graph associated
with the contraction. Then  add  on each edge  the  black vertices denoted $v_i^j$, $i=1, \cdots, 2n$ 
and $j=1, \cdots, d$. The resulting graph is neither regular,  nor properly edge-colored. 
It is however bipartite.
We illustrate the contraction of $2n$ tensors in rank 3 by the  diagram
of Figure \ref{coloredgraphs}.  

The number of contractions matches with  the number of permutation triples  
$( \s_1 , \s_2 , \s_3 ) \in   S_{2n} \times {S}_{2n} \times {S}_{2n} $
subjected to the equivalence 
 \bea 
( \s_1 , \s_2 , \s_3 ) \sim ( \, \gamma_1 \s_1  \gamma , \;
\gamma_2\, \s_2 \gamma  , \; \gamma_3 \s_3 \gamma  \,) \,, \quad 
\quad   \gamma_i\in S_n[S_2]\,, 
\g \in S_n\,, 
\label{equiO}
\eea 
where $S_n[S_2]$ is the so-called wreath product of $S_n$ by $S_2$ 
(that is the semi direct product $ S_n \ltimes (S_2)^n$). Its action describes 
as follows: $(S_2)^n$ permutes independently the half-lines of the $n$ $v_i ^j$'s; 
 then, $ S_n $  permutes  the $n$ vertices $v^j_ i$ of a given color $j$, 
 hence the three copies of the wreath products  in rank 3. 

Up to equivalence,  any contraction belongs to the double quotient
\bea
 S_n[S_2]\times S_n[S_2]\times S_n[S_2] \ses  ( {S}_{2n} \times {S}_{2n} \times {S}_{2n} ) / \diag ( {S}_{2n} )\,.
 \label{dcO}
 \eea 
 The computation of the cardinality of double coset requires a  technique different
 from the complex case but it still involves  the Burnside lemma.
 Its last stage makes use of software programs. 
The number of  invariants  in rank $d=3$, that we denote $Z^o_3(2n)$, follows the sequence: 
\bea
\label{eq:z4}
 1; 5; 16; 86; 448; 3580; 34981; 448628; 6854130; 121173330 \,. 
\eea
Note that Read \cite{Read} introduced  orthogonal polynomial techniques to achieve
the same counting. He was able to obtain the first three terms of the sequence in his seminal paper. 

The generalization at any rank $d$ is straightforward. 
Connected tensor invariants can be also generated by the plethystic logarithm. 
At rank 3, one gets
\be
1; 4; 11;  60; 318; 2806; 29359; 396196; 6231794; 112137138\,. 
\ee
Other counting sequences at rank $d=4$ are provided in \cite{ABGD},
and its appendices list Mathematica programs computing both general and connected sequences
at arbitrary $d$. To our knowledge, none of the sequences at $d\ge 4$ are yet reported
in OEIS. 

\subsection{Topological Field Theory TopFT$_2$}
\label{topftO}

TopFT interprets the tensor invariant counting in a different manner. 
Consider the counting of classes in the double coset \eqref{dcO},
 and  the relation \eqref{equiO}. Using Burnside's lemma, one infers that 
\be\label{BL}
Z^o_3(2n)  =  { 1 \over {[n! (2!)^{n}]^3 (2n)!}}  
\sum_{\gamma_i \in S_n[S_2]}\sum_{\s_i \in S_{2n}}
\sum_{\gamma\in S_{2n}} \delta(\gamma_1 \s_1 \gamma \s_1^{-1}) \delta(\gamma_2 \s_2 \gamma \s_2^{-1})
\delta(\gamma_3 \s_3 \gamma  \s_3^{-1})\,, 
\ee
with $\delta$ the Kronecker symbol  on $S_{2n}$. 

\begin{figure}[h]\begin{center}
     \begin{minipage}[t]{.8\textwidth}\centering
      \includegraphics[angle=0, width=10cm, height=3.5cm]{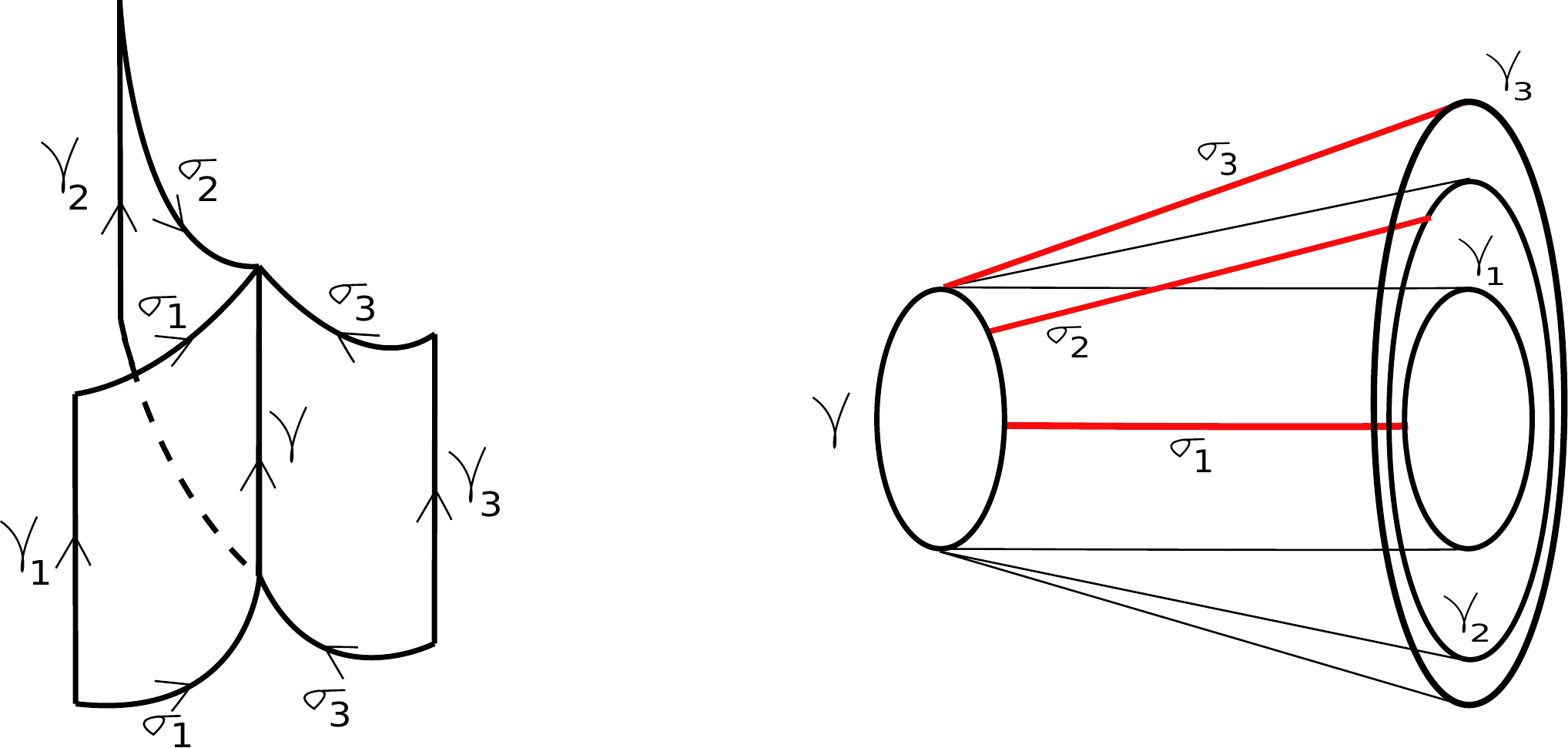}
\caption{ {\small Underlying lattice of TopFT$_2$ related to $Z_3^o(2n)$.}}
\label{tftO}
\end{minipage}
\end{center}
\end{figure}

We identify the above counting as a partition function of a TopFT$_2$ on a cellular complex given 
by Figure \ref{tftO}. Two gauge groups $S_{2n}$ and  $S_n[S_2]$ are needed on this lattice. Associated with that,  three cylinders sharing one of their  
boundary circle characterize the topology of that  2-complex. The enumeration of 3-index orthogonal invariants
corresponds to a $S_{2n}$--TopFT$_2$ on  3 glued cylinders along one circle, with a restriction such that,  the opposite boundary circle is associated with a generator of the gauge group $S_n[S_2]$. 
Such a topological theory has boundary holonomies decorated with $S_n[S_2]$ group elements. 

\begin{figure}[h]
 \centering
     \begin{minipage}[t]{.9\textwidth}
      \centering
\includegraphics[angle=0, width=12cm, height=3.5cm]{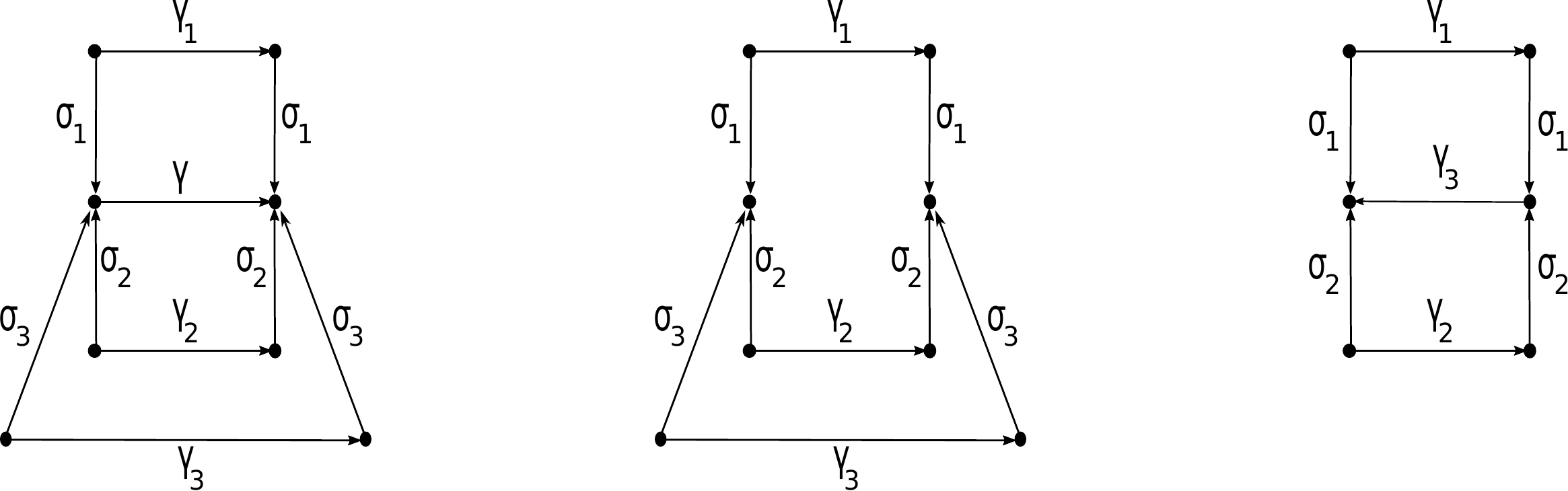}
\caption{ {\small Topological transformations leaving the partition function invariant.}}
\label{opO}
\end{minipage}
\end{figure}

\begin{figure}[h]
 \centering
     \begin{minipage}[t]{.9\textwidth}
      \centering
\includegraphics[angle=0, width=8cm, height=2.5cm]{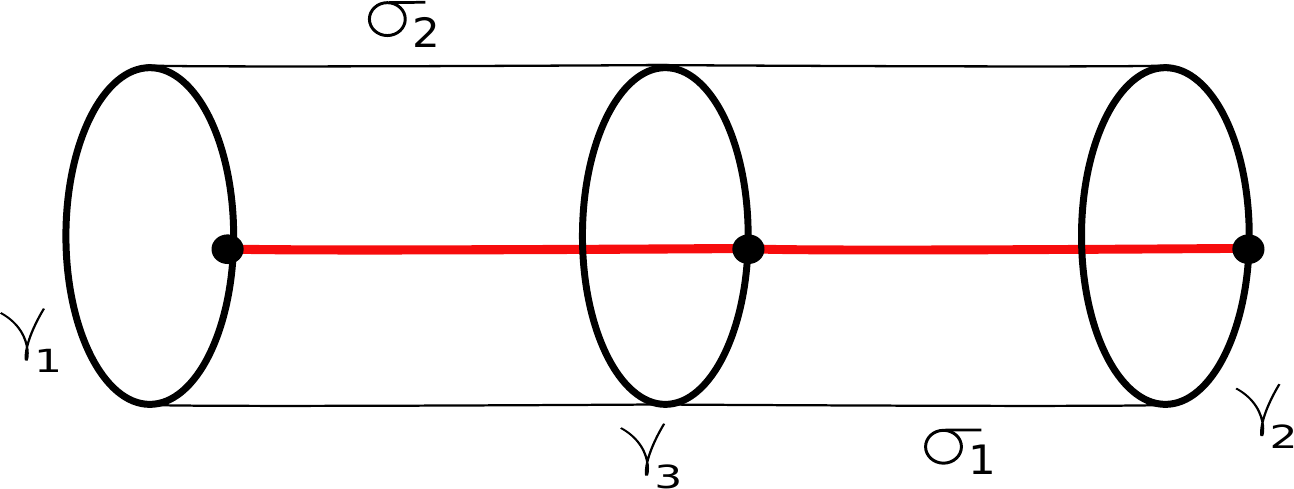}
\caption{ {\small Cylinder with 3 defects.}}
\label{fig:defects}
\end{minipage}
\end{figure}

After a few derivations involving  integration and  changes of variables, we come to 
\be
\label{mass}
Z^o_3(2n) 
= { 1 \over {[n! (2!)^{n}]^3}}  
\sum_{\gamma_i \in S_n[S_2]}\sum_{\s_{1,2} \in S_{2n}}
\delta(\gamma_1 \s_1  \gamma_3 \s_1^{-1}) \delta(\gamma_2 \s_2  \gamma_3 \s_2^{-1}) \, . 
\ee
This integration pictures as in Figure \ref{opO} as the removal of a 1-cell associated
with the variable $\gamma$ in the 2-complex. The partition function therefore shows two types
of invariances: the extraction of $\gamma$ corresponds to one type of topological invariance, and then,
 it is followed by the change of variables $\s_{1,2} \to \s_{1,2} \s_3^{-1} $ corresponding to a  topological invariance 
 of a second kind. 
Thus,  the partition function \eqref{mass} finds another form as
\be
\label{eq:heffect}
Z_3(2n)  =  Z(S^1\times I; (D_{ S_n[S_2] })^{\times 3}) \, ,
\ee
where the right-hand-side defines the partition function obtained by inserting 3 $S_n[S_2]$-defects, one at each end of the cylinder $S^1\times I$, and another one at finite
time $t_0 \in I$, see Figure \ref{fig:defects}. A defect is defined as a closed non-intersecting loop with a marked point. 
The relation \eqref{eq:heffect} shows that orthogonal  invariants 
are in one-to-one correspondence with $n$-fold covers of the cylinder with 3 defects, 
up to  a (symmetry) factor, the stabilizer subgroup of the graph
that we denote ${\rm Aut}(G_{\s_1, \s_2,\s_3})$. 
The order of the stabilizer infers from 
$  {\rm Aut}(G_{\s_1, \s_2,\s_3})  = \sum_{\gamma_i \in S_n[S_2]}
\delta(\gamma_1 \s_1  \gamma_3 \s_1^{-1}) \delta(\gamma_2 \s_2  \gamma_3 \s_2^{-1})
$. This number meets  the number of equivalences
 $(S_n[S_2] \times S_n[S_2] )\ses (S_n \times S_n) / \diag(S_n[S_2]) $ 
 corresponding to a fixed $(\s_1, \s_2)$. 
 
The rank $d$ case tracks the same logic: the counting expresses in terms
of $n$-covers of  $d-1$ cylinders with $d$ defects, one defect shared by all cylinders. It is noteworthy that  TopFT$_2$ may enrich the counting orthogonal invariants with a geometrical picture. 
Indeed, the base space of the TopFT is generally regarded as a string worldsheet \cite{CMR1,CMR2}. 
The counting becomes now counting of worldsheet maps over a cylinder with defects. 
Once more,  this intimates that a link may exist between tensor models and string theory, 
which TopFT could elucidate. This merits full-fledged treatment.

\subsection{Algebras and representation theoretic bases}
\label{reptheorO}
We address the  algebraic structure underlying the counting of real tensor invariants.  
First, we switch to representation theory that  delivers  a different expression
of the same enumeration. Then, this new reading allows us to motivate the search for an algebra
making sense of the new formula. Careful attention must be paid on 
the fact that we might use the same notation as in subsection \ref{reptheor} while
we are dealing with $S_{2n}$ (and not $S_n$). 

\

\noindent{\bf Mapping the counting in representations.}
Let us revisit the  counting \eqref{BL} using the  representation theory of the
symmetric group $S_{2n}$ (the appendices of \cite{ABGD} reviews the main identities used in the following). The  partitions of $2n$ or Young diagrams, $R \vdash 2n$, label
irreps of  $S_{2n}$. 

The counting \eqref{BL} expands as
\bea
\label{eq:KO} 
Z^o_3(2n)  &=&  \frac{1}{[n!(2!)^n]^3(2n)!} \sum_{\gamma_l \in S_n[S_2]} \sum_{\gamma \in S_{2n}} \sum_{R_l \,\vdash\, 2n} \chi^{R_1}(\gamma_1) \chi^{R_1}(\gamma) \chi^{R_2}(\gamma_2) \chi^{R_2}(\gamma) \chi^{R_3}(\gamma_3) \chi^{R_3}(\gamma) \crcr
& = & \frac{1}{[n!(2!)^n]^3}  \sum_{R_l \,\vdash\, 2n} \textbf{C}(R_1,R_2,R_3) \Big[ \sum_{\gamma_1 \in S_n[S_2]} \chi^{R_1}(\gamma_1)\Big]\Big[ \sum_{\gamma_2 \in S_n[S_2]} \chi^{R_2}(\gamma_2)\Big] \, , 
\eea
where the Kronecker coefficient of $S_{2n}$ is defined by
\bea
\textbf{C}(R_1,R_2,R_3) = \frac{1}{(2n)!} \sum_{\gamma \in S_{2n}}\chi^{R_1}(\gamma) \chi^{R_2}(\gamma)
\chi^{R_3}(\gamma) \,. 
\eea
The two other factors $\sum_{\gamma_i \in S_n[S_2]} \chi^{R_i}(\gamma_i)$, $i=1,2$, 
 need a non-obvious treatment. 
To tackle these sums, we use a result by Howe \cite{Howe} (see also \cite{mizukawa,macdo,Ivanov,Caputa:2013hr}): 
$\sum_{\gamma \in S_n[S_2]} \chi^{R}(\gamma)
 = |S_n[S_2]|  \, m^{R}$, where $m^{R}= 1$ if  $R$ is an ``even'' partition, and $m^{R}=0$ otherwise. A partition is called even if all its  row lengths are even. 
Inserting this in \eqref{eq:KO}, we have 
\be\label{kron}
Z^o_3(2n)  =  \sum_{R_l \,\vdash\, 2n/ R_l \text{ is even }} \textbf{C}(R_1,R_2,R_3) \,. 
\ee
A  Sage code implements the sum \eqref{kron} produces the sequence  \eqref{eq:z4} as expected. 

$Z^o_3(2n) $  is also the dimension of an algebra $\cK_{3}(n)$. We will make this clear
in the next developments.  Here again, we realize that counting of colored graphs could 
contribute to the famous problem of giving a combinatorial interpretation to the
Kronecker coefficients \cite{iken1,Blasiak}. From the previous section, it was shown that 
the sum of squares of Kronecker coefficients associated with $S_n$ 
equals the number of $d$-regular bipartite colored graphs made with $n$ black and
$n$ white vertices. Here, the interpretation goes as follows: the number of  $d$-regular 
colored graphs (not necessarily bipartite) equals the sum of all ``even'' Kroneckers
of $S_{2n}$. The two countings are obviously radically different. 
On one hand, the real case associated with $S_{2n}$ have more terms but a linear power of the Kroneckers. 
On the other hand, we may associate the bipartite structure  (or complex tensors) to 
the presence of squares in the sum of Kroneckers. 
Switching from $U(N)$ to $O(N)$, in other words from complex to real, we get rid of the squares but need to deal with a sum  
over a much larger number of terms.  So is there a real Lie group having having even a lesser number
of terms than $O(N)$?

We may finally add that the above counting extends to arbitrary $d$: 
\begin{equation}
Z^o_d(2n)  = \sum_{R_l \,\vdash\, 2n/ R_l \text{ is even } } \textbf{C}_d(R_1,\ldots, R_d) \,, 
 \label{countingZ3n}
\end{equation}
where $\textbf{C}_d(R_1,\ldots, R_d) $ should have now an obvious meaning. 
This expression can be computationally implemented without any issue with Sage.

\

\noindent{\bf $\cK^o_d(n)$, the double coset graph algebra.}
Let us focus on the  algebra generated by real tensor invariants.
We will consider $\mathbb{C}(S_{2n})$, the group algebra of $S_{2n}$. We  first study  the rank $3$ before achieving the general case.

We fix $d=3$. 
Consider  $\sigma_1 \otimes \sigma_2 \otimes \sigma_3$ as an element of the group algebra $\mathbb{C}(S_{2n})^{\otimes 3}$,  and three left actions of 
the subgroup $S_n[S_2]$ and the diagonal right action $\diag(\mathbb{C}(S_{2n}))$ on this triple as:
\begin{equation}
\sigma_1 \otimes \sigma_2 \otimes \sigma_3 \rightarrow \sum_{\gamma_i \in S_n[S_2]} \sum_{\gamma \in S_{2n}} 
\gamma_1 \sigma_1 \gamma \otimes  \gamma_2 \sigma_2 \gamma \otimes  \gamma_3 \sigma_3 \gamma\,.
\end{equation}
$\cK^o_3(2n)$ is the vector subspace of $\mathbb{C}(S_{2n})^{\otimes 3}$ which is invariant under these subgroup actions:
\begin{equation}
\label{Kunalg}
\cK^o_3(2n)= \text{Span}_{\mathbb{C}} \left\{ \sum_{\gamma_i \in S_n[S_2]} \sum_{\gamma \in S_{2n}} \gamma_1 \sigma_1 \gamma \otimes  \gamma_2 \sigma_2 \gamma \otimes  \gamma_3 \sigma_3 \gamma, \quad \sigma_1, \sigma_2, \sigma_3 \in S_{2n} \right\}.
\end{equation}
It is obvious that $\dim \cK^o_3(2n) = Z^o_3(2n)$, since each base element represents a graph equivalence class counted once in $Z^o_3(2n)$. 
Pick two base elements and multiply them
\bea\label{multip}
&&
\Big[\sum_{\gamma_i \in S_n[S_2]} \sum_{\gamma \in S_{2n}} \gamma_1 \sigma_1 \gamma \otimes  \gamma_2 \sigma_2 \gamma \otimes  \gamma_3 \sigma_3 \gamma \Big] \Big[ \sum_{\tau_i \in S_n[S_2]} \sum_{\tau \in S_{2n}} \tau_1 \s'_1 \tau \otimes  \tau_2 \s'_2 \tau \otimes  \tau_3 \sigma_3' \tau \Big] \cr\cr
&&
=\sum_{\tau_i \in S_n[S_2] }\sum_{\gamma \in S_{2n}}  \Big[\sum_{\gamma_i \in S_n[S_2]}
 \sum_{\tau \in S_{2n}} \gamma_1 (\sigma_1 \gamma \tau_1 \s'_1) \tau  \otimes  \gamma_2 (\sigma_2 \gamma \tau_2 \s'_2) \tau \otimes  \gamma_3 (\sigma_3 \gamma\tau_3 \sigma_3' )\tau \Big] . 
\eea
This shows that the multiplication is stable in $\cK^o_3(2n)$, and 
hence,  $\cK^o_3(2n)$ forms graph algebra. 
In the same vein as before,  we can also show that  $\cK^o_3(2n)$ is an associative, unital 
subalgebra of $\mathbb{C}(S_{2n})^{\otimes 3}$ which is semi-simple under a similar
nondegenerate pairing $\delta_3(\cdot,\cdot)$, product of delta's on each factor of $\mathbb{C}(S_{2n})^{\otimes 3}$, see \eqref{delta3}. 
The proof is totally similar for $\cK^o_d(2n)$ (considering $d$ factors in the tensor product)
that is also a semi-simple associative unital subalgebra of $\mathbb{C}(S_{2n})^{\otimes d}$ of dimension $Z^o_d(2n)$.

The semi-simplicity ensures that, by the Wedderburn-Artin theorem,  $\cK^o_d(2n)$  decomposes in matrix subalgebras.  In the complex case and at rank $d=3$, we successfully exhibit the orthogonal base making apparent this matrix decomposition. The real case appears more difficult to handle
and, so far, no Wedderburn-Artin base has been found for $d\ge 3$. 
Postponing this for future investigations, it remains an issue that we could certainly address:
in rank $d=3$, find a representation base with labels  that reflect the dimension \eqref{kron}. This is our  next goal. 

\

\noindent{\bf A representation theoretic base for  $\cK^o_3(2n)$.} 
We start by considering \eqref{fouU} for the present case, and so replace $S_n$ by $S_{2n}$, $\kappa_R$ is a different but fixed constant, 
such that $\{Q^{R}_{ij}\}$ forms this time an orthonormal base of $\C(S_{2n})$: 
  $\delta(Q^{R}_{ij} ; Q^{R'}_{i'j'} )= \delta_{RR'}\delta_{ii'}\delta_{jj'}$. 

Consider the right diagonal action $\rho_R(\cdot )$  and the three left actions $\varrho_i(\cdot )$ on the tensor product $\mathbb{C}[S_{2n}]^{\otimes 3}$. We write:
\begin{equation}\begin{aligned}
& \sum_{\substack{\gamma_1,\,\gamma_2,\,\gamma_3 \in S_n[S_2]}} \sum_{\gamma \in S_{2n}} \varrho_1(\gamma_1)\varrho_2(\gamma_2)\varrho_3(\gamma_3) \rho_R(\gamma) \, Q^{R_1}_{i_1 j_1}\otimes Q^{R_2}_{i_2 j_2}\otimes Q^{R_3}_{i_3 j_3} \\
& = \frac{(2n)!}{d(R_3)} \sum_{\gamma_a} \sum_{p_l \,, q_l}\sum_{\tau} C^{R_1,R_2;R_3,\tau}_{j_1,j_2;j_3}
C^{R_1,R_2;R_3,\tau}_{q_1,q_2;q_3} D^{R_1}_{p_1 i_1}(\gamma_1) 
D^{R_2}_{p_2 i_2}(\gamma_2) D^{R_3}_{p_3 i_3}(\gamma_3)
Q^{R_1}_{p_1 q_1} \otimes Q^{R_2}_{p_2 q_2} \otimes Q^{R_3}_{p_3 q_3}. 
\end{aligned}\end{equation}
Overlapping the result with a Glebsch-Gordan coefficent yields 
\begin{equation}\begin{aligned}
& \sum_{j_l} C^{R_1,R_2;R_3,\tau}_{j_1,j_2;j_3}\sum_{\gamma_a} \sum_{\gamma} \varrho_1(\gamma_1)\varrho_2(\gamma_2)\varrho_3(\gamma_3) \rho_R(\gamma) \, Q^{R_1}_{i_1 j_1}\otimes Q^{R_2}_{i_2 j_2}\otimes Q^{R_3}_{i_3 j_3} 
\label{sumddd}\\
& = (2n)! \sum_{p_l \,, q_l} C^{R_1,R_2;R_3,\tau}_{q_1,q_2;q_3} 
\sum_{\gamma_1} D^{R_1}_{p_1 i_1}(\gamma_1)
\sum_{\gamma_2} D^{R_2}_{p_2 i_2}(\gamma_2)
\sum_{\gamma_3} D^{R_3}_{p_3 i_3}(\gamma_3)
Q^{R_1}_{p_1 q_1} \otimes Q^{R_2}_{p_2 q_2} \otimes Q^{R_3}_{p_3 q_3} \,.
\end{aligned}\end{equation}
An emphasis should be put on  the fact that $\sum_{\gamma \in S_n[S_2]} D^{R}_{p q}(\gamma) \neq 0$,
if and only if $R$ is a partition of $2n$ with even rows. 
Next,   the Wigner matrix element must split using the so-called branching coefficients of  
$S_n[S_2]$ in $S_{2n}$. To proceed with that consider $V^R$ an irreps  $S_{2n}$, and the subgroup inclusion $S_n[S_2] \subset S_{2n}$, we can decompose $V^R$ in irreps  $V^r$  of $S_n[S_2]$ as
$V^R = \oplus_{r} V^{r}\otimes V_{R,r}$, 
where $ V_{R,r}$ is a vector space of dimension the multiplicity of the irreps $r$ in $R$. 
A state in this decomposition is written $\ket{ r, m_r, \nu_r }$, 
where $m_r$ labels the states of $V^{r}$ and $\nu_r = 1, \dots, \dim V_{R,r}$. 
The branching coefficients are defined by the overlap of $\ket{ r, m_r, \nu_r }$ with  an 
orthonormal base of $R$: 
\bea
B^{R;\, r, \nu_r}_{i; \,m_r}  = 
\langle R, i \ket{ r, m_r, \nu_r }  = 
\langle r , m_r, \nu_r\ket{  R, i  }  \, . 
\eea
The properties of $B^{R;\, r, \nu_r}_{i; \,m_r} $ have been listed in \cite{ABGD} (see page 17,18). 
First, decompose the above \eqref{sumddd}, exploiting the branching coefficient
\begin{equation}
\begin{aligned}
& 
\sum_{j_l} C^{R_1,R_2;R_3,\tau}_{j_1,j_2;j_3}\sum_{\gamma_a} \sum_{\gamma} \varrho_1(\gamma_1)\varrho_2(\gamma_2)\varrho_3(\gamma_3) \rho_R(\gamma) \, Q^{R_1}_{i_1 j_1}\otimes Q^{R_2}_{i_2 j_2}\otimes Q^{R_3}_{i_3 j_3} \\
& = (2n)! ( n!2^n) ^3 B^{R_1;tr}_{i_1}B^{R_2;tr}_{i_2}B^{R_3;tr}_{i_3} \sum_{p_l \,, q_l} C^{R_1,R_2;R_3,\tau}_{q_1,q_2;q_3} 
B^{R_1; \, tr}_{p_1} 
B^{R_2; \,  tr}_{p_2} 
B^{R_3;\,  tr}_{p_3} 
Q^{R_1}_{p_1 q_1} \otimes Q^{R_2}_{p_2 q_2} \otimes Q^{R_3}_{p_3 q_3} \,, 
\end{aligned}
\end{equation}
and then we are in position to define the representation base that we are looking for: 
\bea \label{unbasis}
&&
Q^{R_1,R_2,R_3,\tau} 
 = \kappa_{\vec{R}}  
 \sum_{p_l \, ,q_l} C^{R_1,R_2;R_3,\tau}_{q_1,q_2;q_3} 
B^{R_1; \,  tr}_{p_1} 
B^{R_2;\,  tr}_{p_2} 
B^{R_3; \,  tr}_{p_3} 
Q^{R_1}_{p_1 q_1} \otimes Q^{R_2}_{p_2 q_2} \otimes Q^{R_3}_{p_3 q_3} \cr\cr
&& 
=  \kappa_{\vec{R}}  
\frac{\kappa_{R_1}\kappa_{R_2}\kappa_{R_3}}{((2n)!)^3} 
\sum_{\s_i} \sum_{p_l \, ,q_l} C^{R_1,R_2;R_3,\tau}_{q_1,q_2;q_3} 
\Big[\prod_{i=1}^3 B^{R_i; \,  tr}_{p_i} 
D^{R_i}_{p_i q_i}(\s_i)  \Big]
\s_1 \otimes \s_2 \otimes \s_3 \,, 
\label{basis}
\eea
where $\kappa_{\vec{R}}$ is a normalization constant  to be fixed later on, and 
 $\vec R=(R_1, R_2, R_3)$.
Note that  the set $\{Q^{R_1,R_2,R_3,\tau}\}$ is of cardinality 
the counting of  orthogonal invariants \eqref{kron}. 
Some calculations show that  the elements of $\{Q^{R_1,R_2,R_3,\tau}\}$ obey the invariance 
\bea 
(\gamma_1 \otimes \gamma_2 \otimes \gamma_3   )
Q^{R_1,R_2,R_3,\tau}  ( \gamma   \otimes \gamma  \otimes \gamma ) 
  = Q^{R_1,R_2,R_3, \tau}\,, 
\eea
and are orthonomal under the pairing 
\beq \begin{aligned}
 \boldsymbol{\delta}(Q^{R_1,R_2,R_3,\tau}; Q^{R_1',R_2',R_3',\tau'})
= \kappa_{\vec{R}}^2\, d(R_3) 
\sum_{p_l} \Big[ \prod_{i=1}^3 B^{R_i; \,  tr}_{p_i} \Big]^2
\delta_{\vec{R}\vec{R'}} \delta_{\tau \tau'} =  \kappa_{\vec{R}}^2\, d(R_3) \delta_{\vec{R}\vec{R'}} \delta_{\tau \tau'} \, . 
\end{aligned}
 \eeq 
 They form in fact an orthonormal base of $\cK^o_3(2n)$ for a well chosen $\kappa_{\vec R}$. Nevertheless, they do not multiply  like matrices.  Indeed, 
 \beq
Q^{R_1,R_2,R_3,\tau}
Q^{R_1',R_2',R_3',\tau'} = \delta_{\vec{R}\vec{R'}}  k(\vec{R'},\tau) Q^{R_1',R_2',R_3',\tau'} \,, 
\eeq
with a factor $k(\vec{R'},\tau)$ that prevents their orthogonality with respect to the
multiplication. They do not define the base of Wedderburn-Artin matrix decomposition. 
The base $\{Q^{R_1,R_2,R_3,\tau}\}$ decomposes $\cK^o_3(2n)$ in  blocks mutually orthogonal in the labels $(R_1,R_2,R_3)$ but not  in all the remaining labels.

\ 

\noindent{\bf Correlators.} The calculation of correlators happens to be more involved than the  bipartite complexe case. 
In general, correlators build a polynomial in a variable $N$, i.e. the range of a single tensor index, 
with  powers  the number cycles of some permutations. To understand and list these permutations is the first difficult point
to overcome. 

We shall focus on $d=3$ and Wick's theorem allows us to write the 1pt-function of an observable $O_{\bee} = O_{\s_1, \s_2, \s_3} $ as
\bea\label{1ptcorr2}
\la O_{\s_1, \s_2, \s_3}  \ra =  \sum_{\mu \in S_{2n}^*  } N^{ \sum_{i=1}^3 \cy(\mu \ws_i )}\,, 
\eea
where $S_{2n}^*$ is the subset defined by the pairings of $S_{2n}$, that are
permutations  made only of transpositions and
\bea
(\ws_1, \ws_2, \ws_3) = 
(\s_1^{-1} \xi \s_1, \s_2^{-1} \xi  \s_2, \s_3^{-1} \xi \s_3),
\eea 
with $\xi $ the fixed permutation $(12)(34)\dots (2n-1,2n)$.
With the same notation, the normal ordered 2pt-function writes 
\bea\label{2ptcorr}
\la O_{\s_1, \s_2, \s_3}O_{\tau_1, \tau_2, \tau_3}  \ra =  \sum_{\mu \in S_{2n} } N^{ \sum_{i=1}^3 \cy(\mu^{-1}\wt_i \mu \ws_i)}\,. 
\eea
Aiming at finding an orthogonal base of 2pt-functions,  the central elements
 $\Omega_i = \sum_{\alpha_i  \in S_{2n}}  N^{\cy(\alpha_i )-2n} \alpha_i $ will be of great use. 
 We have
 \bea
\label{correl1}
&&
\la O_{\s_1, \s_2, \s_3}O_{\tau_1, \tau_2, \tau_3} \ra 
= 
 N^{6n}
\sum_{\mu  }\delta_3 [ (\mu^{-1})^{\otimes 3} (\wt_1 \otimes  \wt_2 \otimes \wt_3)
 \mu^{\otimes 3} (\ws_1 \otimes  \ws_2 \otimes \ws_3)
 (\Omega_1 \otimes  \Omega_2 \otimes \Omega_3)] \crcr
 &&
 = N^{6n}
\sum_{\mu  } \delta_3  [ (\wt_1 \otimes  \wt_2 \otimes \wt_3)
 \mu^{\otimes 3} (\ws_1 \otimes  \ws_2 \otimes \ws_3)
  (\mu^{-1})^{\otimes 3}
 (\Omega_1 \otimes  \Omega_2 \otimes \Omega_3) ]\,. 
\eea
We then introduce the representation theoretic elements $O^{R_1, R_2, R_3, \tau}  $ as 
\bea
O^{R_1, R_2, R_3, \tau}  
= \sum_{\s_l} \delta_3( Q^{R_1, R_2, R_3, \tau} \s_1^{-1} \otimes \s_2^{-1} \otimes \s_3^{-1} )  O_{\s_1, \s_2, \s_3}
\eea
and, after a lengthy calculation, we arrive at
\bea\label{correlatorRepr}
&& 
\la  O^{R_1, R_2, R_3, \tau}   \, O^{R'_1, R'_2, R'_3, \tau'}   \ra  = 
\Big[  \prod_{i=1}^3 \delta_{R'_i R_i } \Big]\delta_{\tau'\tau}
F(R_1, R_2, R_3,\tau) \crcr
&&
F(R_1, R_2, R_3,\tau) =  \sum_{S_i,\tau_i}
  \Big[ \prod_{i=1}^3   \Dim_N(S_i) \Big]
\Big[\sum_{b_i,c_i,p_i}
D^{S_i}_{b_ic_i}(\xi) 
C^{S_i,S_i;R_i,\tau_i}_{b_i,c_i;p_i}
B^{R_i; \,  tr}_{p_i}
\Big] ^2 \,. 
\eea
This is precisely the orthogonality of the representation theoretic base $\{O^{R_1, R_2, R_3, \tau} \}$ 
for normal ordered Gaussian correlators in $\cK^o_3(2n)$.

%----------------------------------------------------------------------------------------
\section{Conclusion}
\label{concl}
We have enumerated rank $d$ real and complex tensor invariants that are orthogonal
and unitary invariants relying on techniques build on symmetric groups  and their representation theory. 
From this enumeration, we find several bridges with other formalisms. 
In particular, we were interested in TopFT allowing us to interpret in a different way these countings 
as the number of covers of topological objects (punctured 2-sphere for the complex case and torus with defects for the real case) and even to foresee geometrical pictures attached to them. If this program is successfully achieved  then one could establish, for instance, connections between string theory  and tensor models. Such correspondences deserve attention. From another side, the representation theory of the symmetric group brings a different perspective on these countings and
might  connect them with 
the problem of the combinatorial interpretation of the  Kronecker coefficient. 
This could be of major interest in  computational complexity theory. 
Moreover, we have found that tensor model observables span a graph algebra with interesting properties
such as associativity and semi-simplicity. With the latter feature, the complex case exhibits a Wedderburn-Artin base. Finally, we were interested in Gaussian correlators, showing that
there are representation theoretic orthogonal bases for the normal ordered 2pt-function. 
Aiming at extracting physical properties, more work must be performed on computable sectors in tensor models. With the help of computer softwares, the fast calculation of the 1pt-function is 
encouraging. Higher order correlators need to be addressed after this. 

The following table delivers a summary of the  results presented in this work.

\begin{center}
\begin{tabular}{ |c||c|c| }
       \hline
                                        & Unitary TM & Orthogonal TM\\
                                        \hline 
                                        & & \\
Counting observables $d$  &    $\sqrt{} $  &    $\sqrt{} $   \\
 \# of observables at $d=3$  &    1; 4; 11; 43; 161;  \dots &     1; 5; 16; 86; 448; \dots   \\
TopFT$_2$  interpretation                        &    Branched covers of the 2-sphere  & Covers of 2-torus with  defects \\
Algebraic structure       & associative unital semi-simple & associative unitary semi-simple \\
Invariant ortho. rep. base & $\sqrt{} $      &  $\sqrt {}$        \\
1-pt and 2-pt correlators             &$\sqrt{} $  &$\sqrt{} $ \\ 
Wedderburn-Artin base & $\sqrt{} $ & X \\
\hline 
\end{tabular}
\end{center}

Finally, we may venture a few words about new tensor models mixing both $U(N)$ and $O(N)$
invariance. Inspired by the work of Ferrari \cite{Ferrari:2017ryl},  
a tensor model with mixed invariance under $U(N)^{\times 2}\times O(N)$ has been recently studied
by Benedetti et al  \cite{Benedetti:2020iyz}. Note that this type of mixed invariance 
was first introduced in the so-called multi-orientable tensor model  \cite{Tanasa:2015uhr}.
Both these mixed-type models admit a large $N$ expansion and have interesting scaling properties. 
We can certainly apply the above counting formalism to such 
models promoting a mixed group $U(N)^{\times p}\times O(N)^{\times q}$. 
For simplicity, consider the case of a rigid tensor (no symmetry under the tensor index), 
of rank $d = p+q$, require that the number  of $T$ in the observable  is even
(same for the number of $\bar T$), and that no indices associated with an orthogonal invariance
 contract  a $T$ and a $\bar T$  (apart  from the quadratic invariant $\Tr_2(T\bar T)$). 
The counting operates with three groups of permutation 
$(\s_1, \s_2, \dots, \s_p) \in S_{2n}^{\times p}$ implementing the connection between $T$'s and $\bar T$'s,  $(\tau_1, \tau_2, \dots, \tau_q)\in S_{2n}^{\times q}$ establishing the connection between the $T$'s only, and  $(\bar\tau_1,\bar \tau_2, \dots, \bar\tau_q)\in S_{2n}^{\times q}$  connecting the $\bar T$'s between themselves. The invariants are fully determined by the equivalence
\begin{align}
&
[ \tau_1, \tau_2, \dots, \tau_q ;\;\;
\s_1, \s_2, \dots, \s_p;\;\;
\bar\tau_1,\bar \tau_2, \dots, \bar\tau_q]  \crcr
&
\sim 
[ \g_1 \tau_1  \mu_1, \g_2 \tau_2 \mu_1, \dots, \g_q \tau_q  \mu_1;\;\;
\mu_1 \s_1 \mu_2 ,  \mu_1 \s_2 \mu_2, \dots, \mu_1 \s_p \mu_2; \;\;
\mu_2 \bar\tau_1 \bar\g_1  , \mu_2 \bar\tau_2 \bar\g_2 , \dots, \mu_2 \bar\tau_q \bar\g_q]  \crcr
&
\end{align}
where $\g_i$ and $\bar \g_i \in S_{2n}[S_2]$, and $\mu_i \in S_{2n}$. Note that this is no longer a pure
left and then right action as in the cases treated above. As preliminary thoughts, sorting these classes may require to 
embed it in tensor product spaces: $\C(S_{2n} ^{\times q}) \otimes \C( S_{2n} ^{\times p}) \otimes \C(S_{2n} ^{\times q}) $. Then define, four actions: $S_{2n}[S_2]^{\times q }\otimes 1\otimes 1 $ that acts
on the left and only on the first slot;   ${\diag}_{ q\otimes p }^{R,L} ( {S}_{2n} )\otimes 1$ acts by diagonal (right, left) multiplication: the 
right multiplication acts on the $q$ $\tau_i$'s on the first slot, and  the left multiplication 
on the next $p$ $\s_i$'s factors on the second slot; next we have  $1\otimes {\diag}_{ p\otimes q }^{R,L} ( {S}_{2n} )$, that acts similarly (note the exchange of roles
of $p$ and $q$) but on the second and third slots;  finally 
$ 1\otimes 1\otimes  S_{2n}[S_2]^{\times q } $ acting on the right and only on the third slot: 
\bea
&&
[S_{2n}[S_2]^{\times q }\otimes 1\otimes 1 ]
[{\diag}_{ q\otimes p }^{R,L} ( {S}_{2n} )\otimes 1]
[1\otimes {\diag}_{ p\otimes q }^{R,L} ( {S}_{2n} )]
[1\otimes 1\otimes  S_{2n}[S_2]^{\times q }]  \crcr
&&
\qquad 
\triangleright \; 
\C(S_{2n} ^{\times q}) \otimes \C( S_{2n} ^{\times p}) \otimes \C(S_{2n} ^{\times q}) \;. 
\eea
The fact that ${\diag}_{ q\otimes p }^{R,L} ( {S}_{2n} )\otimes 1$ and 
$1\otimes {\diag}_{ p\otimes q }^{R,L} ( {S}_{2n} )$ do not form groups make 
this new puzzle interesting.

\section*{Acknowledgments}
I gratefully thanks the organisers of Corfu Summer Institute 2019 ``School and Workshops on Elementary Particle Physics and Gravity'' (CORFU2019) and the Humboldt Kolleg ``Frontiers in Physics, 
From the Electroweak to the Planck Scales'', especially George Zoupanos, Patrizia Vitale and Ifigeneia Moraiti. 
I am also particularly indebted to Sanjaye Ramgoolam for introducing me in the subject and for years of collaboration without which this work could not stand. I wish also to warmly thank Avohou Remi Cocou and Nicolas Dub, for collaboration on the joint work presented here.

\end{document}